# Lateral forces on nanoparticles near a surface under circularly-polarized plane-wave illumination

Francisco J. Rodríguez-Fortuño[1], Nader Engheta[2], Alejandro Martínez[3] and Anatoly V. Zayats[1]*

[1]King's College London, Strand, London WC2R 2LS, United Kingdom
[2]University of Pennsylvania, Philadelphia, Pennsylvania 19104, USA
[3]Nanophotonics Technology Center, Universitat Politècnica de València, Valencia 46022, Spain

* Corresponding author: a.zayats@kcl.ac.uk

**Abstract.** Optical forces allow manipulation of small particles and control of nanophotonic structures with light beams. Here, we describe a counter-intuitive lateral optical force acting on particles placed above a substrate, under uniform plane wave illumination without any field gradients. We show that under circularly-polarized illumination, nanoparticles experience a lateral force as a result of dipolar, spin-sensitive scattering, with a magnitude comparable to other optical forces. To this end, we rigorously calculate the force experienced by a circularly polarized dipole radiating above a surface. Unlike for linearly-polarized dipoles, force components parallel to the surface can exist, caused by the recoil of unidirectional guided modes excited at the surface and/or by dipole-dipole interactions with the induced image dipole. These results were presented and discussed in conferences [1] and [2].

The transfer of momentum from electromagnetic radiation to matter enables the existence of optical forces [3,4], allowing the trapping and manipulation of particles [5–8], single cells [9], and atoms [10] with important applications such as optical cooling [11] or lab-on-a-chip microfluidic sorting [12] amongst many others [13]. There is a wide variety of optical forces, such as radiation pressure force [5,6], exotic metamaterial forces for adhesion or repulsion [14–17], and gradient forces in optical tweezers [13,18], waveguides [19–21] and nanostructures [22,23]. Yet, all these examples are either (i) localized to a single focused beam (e.g. optical tweezers) or field gradient in the vicinity of a fixed nanostructure, or (ii) allow limited manipulation, with the force acting in the illumination direction (e.g. pressure force) or perpendicular to a substrate. The possibility of a force that acts simultaneously on several particles at different locations within a wide area (not requiring focusing of a light beam onto the individual objects), and, in addition, is directed laterally (parallel to the substrate and perpendicular to the illumination direction), would enable the mass movement, arrangement and sorting of particles on a substrate or waveguide in a simple way. Recent theoretical proposals have unveiled a lateral optical force by using chiral particles near a substrate [24] or unusual lateral momentum carried by evanescent waves near a surface where particles are placed [25,26]. However, while the former works only on chiral particles, the latter requires near-field illumination. We ask whether lateral forces, normal to the plane of incidence, can be achieved for arbitrary objects with far-field, plane wave illumination, acting on an ensemble of objects simultaneously.

The nature of a point dipole and the rotational symmetry of a plane surface intuitively suggest that the total time-averaged force acting on the dipole will always be aligned on an axis perpendicular to the surface, either attracting the dipole towards the surface or repelling it away. Indeed, this is the case for linearly polarized dipoles, for which the force –always directed along the normal to a surface– has been widely studied [17,27], being the origin of, e.g., the van der Waals and Casimir attractive forces [3,28,29]. Only when symmetry is broken, such as by using a chiral particle [24], a lateral force oriented along the surface can appear. Recently, however, it was shown that symmetry can also be broken by using circularly polarized dipoles (or circularly polarized exciting electromagnetic fields) [30–33]. Such spin-carrying dipoles, placed in near-field proximity to a surface, lead to a preferred direction of excitation of guided and radiated electromagnetic fields [30–48] which, following conservation of momentum, suggests the intriguing possibility of a lateral force acting on them. This would imply that the dipolar scattering of polarizable particles placed on a substrate, driven by a circularly polarized plane wave illuminating a wide area, could exert a lateral force on the particles [Fig. 1(a)], with a polarization-controlled switching of its direction.

In this manuscript we demonstrate, analytically and numerically, the presence of such counterintuitive lateral force, which does not exist for linearly polarized dipoles, and which switches its direction for circularly polarized dipoles of opposite handedness, thus providing opportunity to control it uniquely with the polarisation of the illuminating light. We identify and quantitatively describe different contributions to this force: (i) if the nearby surface supports guided modes (plasmonic or photonic), there is a recoil-type force due to their unidirectional excitation; (ii) even in the absence of guided mode excitation, we found an additional lateral force caused by the $\pi/2$ out-of-phase component of the circularly-polarized image dipole induced in the lossy substrate. Whilst our model considers a dipole radiating near a surface, it is directly applicable to the scattering of illuminated polarizable particles [30,49], yet it is not limited to that particular situation.

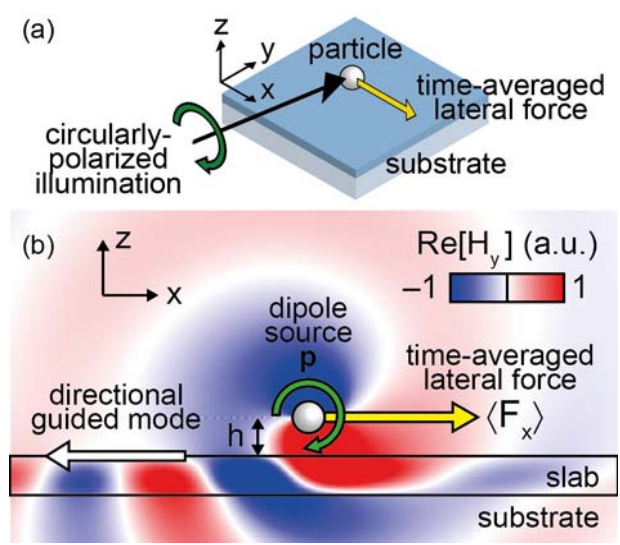

Figure 1. (a) Schematic of far-field illumination lateral force. (b) Dipole scattering: simulated magnetic field of a circularly polarized dipole at a distance h above a dielectric slab. The dipole handedness (green), the direction of the guided mode excitation in a slab (white), and the lateral force (yellow) are indicated by coloured arrows. The parameters of the model correspond to example 1 in the text.

We consider the general case [Fig. 1(b)] of a point dipole source placed in free space at a height $h$ above a planar surface at $z = 0$. From the point of view of the dipole, the surface is completely characterised by its Fresnel reflection coefficients $r^p(k_t)$ and $r^s(k_t)$ for p-polarized and s-polarized waves, respectively, where $k_t = (k_x^2 + k_y^2)^{1/2}$ is the transverse wave-vector. This formulation is valid for a substrate, a slab, and any other collection of stacked slabs of different materials. Assuming a time-harmonic scenario with angular frequency $\omega = 2\pi c_0 / \lambda$, we can work with complex phasor notation, so that the polarization of the dipole is $\mathbf{p}(t) = \text{Re}\{\mathbf{p} \exp(-i\omega t)\}$, where $\mathbf{p} = [p_x, p_y, p_z]$. We derive the force from first principles, considering the dipole as oscillating equal charges with opposite signs $\pm q$, which radiate an electromagnetic field. Upon interaction with the substrate, the radiated field will be reflected back and exert an electromagnetic force on the charges, in accordance to $\mathbf{F} = q(\mathbf{E} + \mathbf{v} \times \mathbf{B})$. When this equation is applied to the two oscillating charges of a single point dipole and the resulting force is averaged in time [3], it can be written in terms of the electric field only, as $\langle \mathbf{F} \rangle = (1/2)\text{Re}\{p_x^* \nabla E_x + p_y^* \nabla E_y + p_z^* \nabla E_z\}$, where $\nabla = [\partial/\partial x, \partial/\partial y, \partial/\partial z]$ is the gradient operator, evaluated at the location of the dipole. Since the aim of this manuscript is to study the lateral forces (perpendicular to the z-axis) acting on the dipole, we will focus on the x-component of the force, without loss of generality, since the x-axis can be reoriented into any desired direction in the x-y plane. The calculations are greatly simplified by working with the spatial frequency decomposition of the fields in the $k_x$-$k_y$ plane, known to be asymmetric for circularly polarized dipoles [30]. After substituting the fields of the dipole reflected by the surface into the equation of the force (see supplemental materials SM), we arrive at a compact exact equation for the time averaged lateral force acting on the dipole:

$$\langle F_x \rangle = -\frac{3 P_{rad}^{xz}}{4 c_0} \eta_{pol} \int_0^{+\infty} k_{tr}^3 \, \text{Im}\left\{ r^p(k_{tr}) \cdot e^{i2\pi \frac{2h}{\lambda}\sqrt{1-k_{tr}^2}} \right\} dk_{tr} \qquad (1)$$

where the integration is performed over the normalized transverse wave-vector variable $k_{tr} = k_t / k_0$, $k_0 = \omega/c$, $P_{rad}^{xz} = \omega^4 [|p_x|^2 + |p_z|^2]/[12\pi\varepsilon_0 c_0^3]$ is the power radiated by the $x$ and $z$ components of the dipole if it was placed in free space, used as an experimentally relevant measure of the amplitude of the dipole excitation, and $\eta_{pol} = 2\,\text{Im}\{p_x p_z^*\}/[|p_x|^2 + |p_z|^2] = [|p_{lcp}|^2 - |p_{rcp}|^2]/[|p_{lcp}|^2 + |p_{rcp}|^2]$ is a polarization chirality factor which equals $\pm 1$ for right and left circularly polarized dipoles, respectively, and 0 for linearly polarized dipoles (which therefore do not experience any lateral force). The form of $\eta_{pol}$ is identical to the third Stokes parameter of a polarized electric field, often used as a measure of the local "chirality" or "spin" carried by the fields. This chirality in the polarization of the dipole is ultimately responsible for the unidirectional excitation of guided modes and the appearance of a lateral force. Notice that the equation is independent of frequency (for a fixed $h/\lambda$). Only the *p*-polarized components of the field appear in Eq. (1) and, therefore, only transverse magnetic (TM) modes excited in the surface will affect the force. Also notice that the *y*-component of the electric dipole does not affect the *x*-component of the lateral force.

For added physical insight, it is highly convenient to approximate the force (Eq. 1) as three separate contributions (see SM):

$$\langle F_x \rangle \approx \sum_{k=1}^{N} \langle F_x \rangle_{\substack{\text{recoil}\\\text{mode k}}} + \langle F_x \rangle_{\substack{\text{image}\\\text{dipole}}} + \langle F_x \rangle_{\substack{\text{propagating}\\\text{components}}} \quad (2)$$

The accuracy of the approximation (Eq. 2) and the distinct behaviour of the different terms is clearly observed in Fig. 2, showing the distance *h* dependence of the total force together with the contribution from the different terms.

The first term is the "recoil" force caused by the directional excitation of any TM modes supported by the surface, labelled as $k = 1, \ldots, N$. If the surface is metallic, a single surface plasmon mode exists (*N*=1). A dielectric slab, on the other hand, might support several TM guided modes. In the absence of guided modes, such as in a dielectric substrate, N=0, and the first term vanishes. It is evident, as depicted schematically in Fig 1(b), that the unidirectional excitation of momentum-carrying electromagnetic modes by a spin-carrying dipole implies –in accordance with the conservation of momentum– the existence of a "recoil" force acting on the source (like the recoil experienced by atoms upon emission of photons [50]). This lateral force is directed opposite to the mode excitation and therefore, parallel to the surface. A simple expression for this term can be deduced by recognizing that any mode will manifest itself in Eq. (1) as a sharp resonant peak in the reflection coefficient $r^p$. We can approximate the imaginary part of $r^p$ in Eq. (1) with a sum of Dirac δ-functions corresponding to the different modes $\text{Im}(r^p) \approx \sum_k R_k \delta(k_{tr} - n_{\text{eff},k})$, where $n_{\text{eff},k} > 1$ is the effective mode index of the *k*-th mode, and $R_k = \int_{n_{\text{eff},k}-\Delta_k/2}^{n_{\text{eff},k}+\Delta_k/2} \text{Im}(r^p) dk_{tr}$ is a measure of the total reflected amplitude of the mode (proportional to the amplitude of the mode excitation), with $\Delta_k$ being a bandwidth of integration that embraces the whole resonant peak. After the integration of Eq. (1) the recoil force can be obtained as:

$$\langle F_x \rangle_{\substack{\text{recoil}\\\text{mode k}}} \approx -\frac{3 P_{rad}^{xz}}{4 c_0} \eta_{pol} n_{\text{eff},k}^3 R_k e^{-2\pi \frac{2h}{\lambda}\sqrt{n_{\text{eff},k}^2 - 1}} \quad (3)$$

This equation carries major physical insights (green lines in Fig. 2): The force is bigger for higher mode index $n_{\text{eff},k}$ and excitation amplitude $R_k$, as expected due to the increased momentum and amplitude, respectively, of guided photonic or plasmonic modes. The distance dependence accounts for the evanescent decay of near-fields for a length 2*h*, corresponding to the "round-trip" of the dipolar near-fields exciting the surface mode, which in turn acts back on the dipole. The sign of the polarization factor $\eta_{pol}$ is given by the rotation sense of the dipole, which determines the mode excitation direction [30] and, therefore, the direction of the recoil force.

The second term in Eq. (2) corresponds to dipole-dipole interactions under the image-dipole quasistatic approximation, accounting for the non-resonant parts of the integrand of Eq. (1) at high wave vectors values

$k_{tr} \gg 1$. A simple expression for this term can be obtained by assuming that $r^p$ is constant and equal to its quasistatic value $S = \lim_{k_t/k_0 \to \infty} r^p = (\varepsilon_2 - 1)/(\varepsilon_2 + 1)$, where $\varepsilon_2$ is the permittivity of the first material slab at $z = 0$. The approximation $r^p(k_{tr}) \approx S$ corresponds to quasistatic image theory, in which every charge $q$ has an associated mirror image $q' = -qS$. The image dipole is thus instantaneously correlated to the source by the complex amplitude $S$. This simplification makes Eq. (1) analytically solvable, resulting in (see SM):

$$\langle F_x \rangle_{\substack{\text{image} \\ \text{dipole}}} \approx -\frac{3 P_{rad}^{xz}}{4 c_0} \eta_{pol} \frac{3}{128 \pi^4} \text{Im}(S) \cdot \left(\frac{h}{\lambda}\right)^{-4} \tag{4}$$

This force, having an $h^{-4}$ dependence, dominates over the other components at the limit of very small distances and exists even in the absence of guided modes in the substrate or slab. The same analytical expression can also be deduced by calculating the force between two dipoles (see SM) related by the image coefficient $S$, placed at a distance $2h \ll \lambda$. Intriguingly, this tells us that, in general, a dipole exerts a force on another nearby dipole in a direction different to that of the line joining their centres. This finding is certainly surprising given the two-point nature of such geometry. For the lateral force to exist, the condition $|\eta_{pol}| > 0$ needs to be satisfied, i.e. the dipoles must have a chirality in their polarization in order to break the *x*-mirror symmetry. In addition, the term $\text{Im}(S)$ means that only those dipole components which are $\arg(S) = \pi/2$ out of phase with the other dipole contribute to this force, i.e., the polarization of the image dipole must lag behind in order to produce a lateral force. This means that a lateral force between a dipole and its image necessarily requires a lossy substrate in order to provide the required phase difference. In lossless materials, the image dipole is always in phase $\arg(S) = 0$ or anti-phase $\arg(S) = \pi$ to the source dipole, and the lateral force vanishes. This is in stark contrast to the vertical attractive or repulsive force from a substrate due to the image dipole [17,27], which depends on $\text{Re}(S)$. Those two very particular conditions (circular polarization and losses) contribute to the elusive nature of this force in previous studies.

Finally, the third term in Eq. (2) is the force caused by the reflected propagating components of the dipole, given by the integral of Eq. (1) with its limits modified to $k_{tr} \in [0,1]$, thus neglecting the evanescent components ($k_{tr} > 1$) of the fields. In the proximity of the substrate, this term is overshadowed by the other two terms, but it is the only surviving term in the far field (Fig. 2). This term reverses its sign with $h$ due to phase advance.

We now provide two numerical examples to illustrate these forces. First, the very general scenario of a circular dipole over a dielectric slab (Fig. 1), valid for any wavelength λ for which the dielectric is lossless. This can represent a non-resonant spherical nanoparticle illuminated by circularly polarized light, in the vicinity of a dielectric slab (in this case, silicon, $n = 3.45$) on top of an infinite substrate (silica $n = 1.45$). The thickness of the slab $t = 0.135 \lambda$ was optimized to achieve a maximum force (see SM, Fig. S1). Fig 2(a) depicts the lateral force acting on the dipole $\langle F_x \rangle$ per unit radiated power $P_{rad}^{xy}$. Since we assume that the dielectric slab is lossless, the image dipole contribution to the force [Eq. (4)] vanishes. The slab however supports two TM guided modes ( $n_{\text{eff},1} = 1.780, R_1 = 0.822$; $n_{\text{eff},2} = 1.072, R_2 = 0.286$ ), which will be unidirectionally excited by the dipole, and will exert a recoil force on the particle according to Eq. (3). Each mode shows its own exponential decay with the distance $h$: Mode 1 decays faster due to its higher effective index, but exerts a greater force as $h \to 0$ due to the higher momentum and amplitude of excitation.

Our second example is a circularly polarized dipole above a lossy gold substrate, modelled with a relative permittivity $\varepsilon_2 = -11.796 + 1.2278i$ at the wavelength of $\lambda = 632.8$ nm. Such a metal surface sustains the unidirectional excitation of a surface plasmon polariton mode ( $n_{\text{eff,SPP}} = 1.045, R_{\text{SPP}} = 0.596$ ) with the associated lateral recoil force. As expected [Eq. (3)], the strength of the recoil force exponentially decays away from the surface [Fig. 2(b)]. At small distances from the surface, the main contribution to the force comes in this case from the image dipole contribution. Due to the imaginary part of the Au permittivity, the image dipole has the required phase-lag $\arg(S)$ with respect to the source dipole and exerts a lateral force on it with an $h^{-4}$ distance dependence [Eq. (4)]. To cross-check the results, Fig. 2(b) includes also the force obtained by evaluating numerically the spatial derivatives of the electric field using Green's function formalism [51], showing an exact correspondence with Eq. (1). The time-domain numerical simulations of the fields, and a subsequent calculation of the force by integrating the Maxwell's stress tensor in a box around the particle, also provides excellent agreement (SM, Fig. S4).

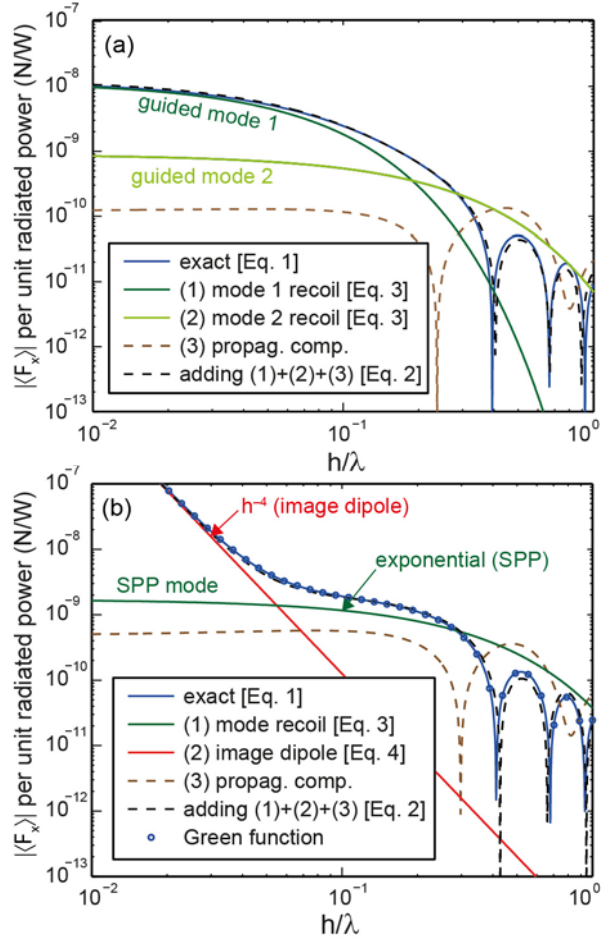

FIG 2. Distance dependence of the time-averaged lateral force for a circularly polarized dipole p = [1, 0, i] above (a) a dielectric slab (n=3.45) of thickness t = 0.135λ resting on a substrate (n=1.45) and (b) a gold substrate ($\varepsilon_2$ = -11.796 + 1.2278i at 632.8 nm). The exact result from Eq. 1 and the different terms in the approximation are shown. Note: For the calculation of $R_k$ we used $\Delta_1$=0.3$n_1$, $\Delta_2$=0.15$n_2$ and $\Delta_{SPP}$=0.1$n_{eff,SPP}$. The results of the Green function evaluation are also shown in (b). Results based on the Maxwell's stress tensor calculations are shown in Fig. S4.

Both examples illustrate the wide applicability of such lateral forces. In the above discussion, we have calculated the normalized force $\langle F_x \rangle / P_{rad}^{xz}$. However, in order to provide practical insight, it would be highly desirable to evaluate the order of magnitude of the force in experimental situations. To do this, we have to estimate the power radiated by the dipole $P_{rad}^{xz}$, which can be a quantum dot or an optical, microwave or radio-frequency antenna, so that the radiated power is determined by the source driving the dipole. Alternatively we can consider a simpler optical experiment [Fig. 1(a)]: a polarizable nanoparticle (isotropic polarizability $\alpha$) being illuminated by an external light source, at a grazing angle so that it gets polarized with a polarization $\mathbf{p} = \alpha\mathbf{E}_{\mathbf{inc}}$ in the x-z plane. The particle will re-radiate part of the incident power as dipolar scattering ($P_{rad}^{xz}$ can be approximated to this scattering $P_{sca} = \sigma_{sca}W_{inc}$, determined by the product of the scattering cross section of the particle and the incident power density). A small spherical gold particle [52] of radius $R = 30$ nm has a scattering cross section in the dipole/Rayleigh approximation (see SM, Fig. S5) of 1100 nm$^2$ at its localized plasmon resonance $\lambda = 520$ nm. For a typical 10 mW laser illuminating a spot of radius 10 μm, the scattered power by the gold nanoparticle is 3.5×10$^{-8}$ W. This value multiplies the normalized force graphs (Fig. 2), resulting in lateral forces of the order of 10$^{-16}$ to 10$^{-17}$ N, comparable with the weight of the particle W = 2.14×10$^{-17}$ N. The illumination will also give rise to other forces, such as the transfer of linear momentum to the particle, which exerts a pressure force [3] equal to 3.3×10$^{-16}$ N along the illumination direction. Therefore, although the lateral force is not necessarily the dominant force, it clearly is comparable with other electromagnetic forces and is suited to experimentation and practical applications. The fact that the lateral force acts orthogonally to other forces [Fig. 1(a)] in a practical situation (including pressure and gradient forces), together with its polarization dependence, makes it susceptible to a relatively easy experimental discrimination.

In conclusion, we have explored the physical origins of a lateral force acting on a dipole placed in near-field proximity to a surface or planar waveguide, which can be induced by plane wave illumination in the direction normal to it and not requiring field gradients. In the case of an Au nanoparticle illuminated at its plasmon resonance wavelength, the magnitude of such force is comparable with the particle's weight and with other acting electromagnetic forces. Studies of optical forces on small particles usually ignore the scattering of the particle in comparison to the absorption, overlooking the complex forces caused by the former. The scattering of a non-chiral particle over a substrate allows a lateral force to exist, even if neither the particle nor the surface break the left-right symmetry, by using circular polarization in the illuminating plane wave. The polarization itself breaks the symmetry, and the resulting force depends directly on its "chirality", allowing a precise and broadband optical control of its direction and magnitude.

The proposed effect is not limited to a particular realisation, because it involves the lateral force on a polarized dipole, which can be achieved throughout the electromagnetic spectrum using a variety of materials. It can be used to model the scattering of a nanoparticle illuminated with circularly polarized light, but it can also model other scenarios, e.g., linearly polarized illumination of tilted ellipsoidal particles achieving circular polarization [48], or circularly polarized RF antennas. It is obvious that a lateral force will also exist in non-planar geometries, such as dipoles near a cylindrical waveguide [41,47], and in general any geometry where spin-orbit interaction can be exploited [43,46] to achieve spin-controlled recoil forces. The possible implications of a controllable lateral force existing between a dipole and a nearby surface (or between two dipoles) are wide-spanning from microfluidics to optomechanics, reconfigurable antennas and metamaterials.

**Acknowledgements**

This work has been supported, in part, by EPSRC (UK). AZ acknowledges support from the Royal Society and the Wolfson Foundation. NE acknowledges partial support from the US Office of Naval Research (ONR) Multidisciplinary University Research Initiative (MURI) Grant No. N00014-10-1-0942. AM acknowledges support from the Spanish Government (contract No. TEC2011-28664-C02-02).

# Supplemental Materials

- 15 pages of supplemental text: S1–S15.

- 4 Supplemental Material Figures: Figs. S1–S5.

## Supplemental Material contents:

# 1. Derivation of the exact equation for the lateral force

Consider a dipole $\mathbf{p} = [p_x, p_y, p_z]^T$ at a distance $h$ above a planar surface (plane $z = 0$). We will consider the general case in which the upper medium has a relative permittivity $\epsilon_1$ and relative permeability $\mu_1$, while in the main text both values are taken as 1 corresponding to free space. The surface can be a semi-infinite substrate, or any number of stacked slabs, but from the point of view of the dipole, the surface is *entirely characterised* by the Fresnel reflection coefficients $r^p(k_t)$ and $r^s(k_t)$ for p-polarized and s-polarized fields, respectively, where $\mathbf{k_t} = k_x\hat{\mathbf{x}} + k_x\hat{\mathbf{y}}$ is the transverse wave-vector associated with a given plane wave. Values of $|k_t| > k_1$ represent evanescent components, where $k_1 = n_1 k_0 = (\epsilon_1 \mu_1)^{1/2} k_0$ and $k_0 = 2\pi/\lambda$. In this section we will derive the time-averaged lateral component of the force acting on a dipole due to its own fields reflected by the surface.

The electromagnetic force acting on an ideal dipole due to external fields $\mathbf{E}$ and $\mathbf{B}$ can be calculated from first principles using the Lorentz electromagnetic force acting on each of the two point charges $\mathbf{F} = q(\mathbf{E} + \mathbf{v} \times \mathbf{B})$ that constitute the dipole, and written in terms of the spatial derivatives of the electric field at the location of the dipole. For time-harmonic fields of angular frequency $\omega$, the time-averaged electromagnetic force on the dipole is given as (see Ref. [1]):

$$\langle \mathbf{F} \rangle = \frac{1}{2} \operatorname{Re} \left[ p_x^* \nabla E_x(\mathbf{r}_{\text{dip}}) + p_y^* \nabla E_y(\mathbf{r}_{\text{dip}}) + p_z^* \nabla E_z(\mathbf{r}_{\text{dip}}) \right], \tag{S1}$$

where the only approximation made was that the dipole is moving slowly compared to the speed of light, and $\nabla = [\partial/\partial x, \partial/\partial y, \partial/\partial z]^T$ is the gradient operator acting on the different components of the electric field $\mathbf{E}(\mathbf{r}) = [E_x, E_y, E_z]^T$ evaluated at the location of the dipole $\mathbf{r}_{\text{dip}} = [0, 0, h]^T$. This electric field corresponds to that applied on the dipole, not including the self-fields of the dipole. In our scenario, $\mathbf{E}(\mathbf{r})$ corresponds to the fields reflected by the surface from the dipole back into itself.

If we are interested in the lateral component of the force, we can focus on the x-component $\langle F_x \rangle$ without loss of generality, since the x-axis can be reoriented into any desired direction within the x-y plane. From eq. (S1) we can write the x-component of the force as:

$$\langle F_x \rangle = \frac{1}{2} \operatorname{Re} \left[ p_x^* \left. \frac{\partial E_x(x,y,z)}{\partial x} \right|_{\substack{x,y=0 \\ z=h}} + p_y^* \left. \frac{\partial E_y(x,y,z)}{\partial x} \right|_{\substack{x,y=0 \\ z=h}} + p_z^* \left. \frac{\partial E_z(x,y,z)}{\partial x} \right|_{\substack{x,y=0 \\ z=h}} \right], \tag{S2}$$

The key idea used in our work is to make use of the spatial decomposition of the electric field, such that it is written as a sum of its plane wave and evanescent components (spatial spectrum):

$$\mathbf{E}(x,y,z) = \iint\limits_{-\infty}^{+\infty} \mathbf{E}(k_x, k_y, z) \cdot e^{ik_x x + ik_y y} \cdot dk_x dk_y, \tag{S3}$$

which can be substituted into eq. (S2), moving the integrals out of the derivatives, and yielding the spatial-spectrum decomposition of the lateral force:

$$\langle F_x \rangle = \iint\limits_{-\infty}^{+\infty} \underbrace{\frac{1}{2} \operatorname{Re} \left[ ik_x p_x^* E_x(k_x, k_y, h) + ik_x p_y^* E_y(k_x, k_y, h) + ik_x p_z^* E_z(k_x, k_y, h) \right]}_{\langle F_x(k_x,k_y) \rangle = (1/2)\operatorname{Re}[ik_x(\mathbf{p}^* \cdot \mathbf{E}(k_x,k_y,h))]} dk_x dk_y, \tag{S4}$$

where $\langle F_x(k_x, k_y) \rangle$ represents the force acting on the dipole caused by each plane wave component of the field, which corresponds to the spatial transform of eq. (S2), where the spatial derivatives $\partial/\partial x$ change to multiplication by $ik_x$. Notice that, although we have been speaking of decomposition into plane waves, the formalism also includes the evanescent waves of the field, which mathematically can be seen as planewaves with a transverse wave-vector greater than the wave-vector in the medium $k_t > k_1$.

Now we only have to substitute the incident fields (reflected by the substrate) into eq. (S4). The analytical form of the reflected fields is developed in Annex I, and the fields are given in eqs. (A4) to (A6b).

**Symmetry simplifications using parity**

For further simplification, notice that eq. (S4) has an integration over $k_x$ and $k_y$ from $-\infty$ to $+\infty$, so that any odd components (odd in $k_x$ or odd in $k_y$) of $\langle F_x(k_x, k_y) \rangle$ will cancel out when performing the integration. Therefore, only the components of $\langle F_x \rangle$ that are *even* with both $k_x$ and $k_y$ will survive the integration, and will be the only ones in which we are interested.

$$\langle F_x \rangle = \iint_{-\infty}^{+\infty} \langle F_x(k_x, k_y) \rangle dk_x dk_y = \iint_{-\infty}^{+\infty} \langle F_x(k_x, k_y) \rangle \Big|_{\substack{k_x \text{ even} \\ k_y \text{ even}}} dk_x dk_y, \tag{S5}$$

By looking at eq. (S4), we can easily deduce that only the components of the field $\mathbf{E}(k_x, k_y, h)$ that are even in $k_y$ *and* odd in $k_x$ will contribute to the force.

$$\langle F_x(k_x, k_y) \rangle \Big|_{\substack{k_x \text{ even} \\ k_y \text{ even}}} = \frac{1}{2} \operatorname{Re} \left[ ik_x \left( \mathbf{p}^* \cdot \mathbf{E}(k_x, k_y, h) \Big|_{\substack{k_x \text{ odd} \\ k_y \text{ even}}} \right) \right] \tag{S6}$$

Thus, we can select and keep only the terms from the reflected field [eqs. (A5) to (A6b) in the Annex A] that show the appropriate parity in $k_x$ and $k_y$, which gives us the following reflected field:

$$\mathbf{E}_{\text{ref}}(k_x, k_y, z) \Big|_{\substack{k_x \text{ odd} \\ k_y \text{ even}}} = \frac{i\omega^2 \mu_0 \mu_1}{8\pi^2} \left( \overset{\leftrightarrow}{\mathbf{M}}^{\mathbf{s}}_{\text{ref}}(k_x, k_y) \Big|_{\substack{k_x \text{ odd} \\ k_y \text{ even}}} + \overset{\leftrightarrow}{\mathbf{M}}^{\mathbf{p}}_{\text{ref}}(k_x, k_y) \Big|_{\substack{k_x \text{ odd} \\ k_y \text{ even}}} \right) \cdot \mathbf{p} \cdot e^{ik_{z_1}(z+h)}, \tag{S7}$$

where the relevant components of the matrices are given by:

$$\overset{\leftrightarrow}{\mathbf{M}}^{\mathbf{s}}_{\text{ref}}(k_x, k_y)) \Big|_{\substack{k_x \text{ odd} \\ k_y \text{ even}}} = 0$$

$$\overset{\leftrightarrow}{\mathbf{M}}^{\mathbf{p}}_{\text{ref}}(k_x, k_y)) \Big|_{\substack{k_x \text{ odd} \\ k_y \text{ even}}} = \frac{-k_x r^p(k_x, k_y)}{k_1^2} \begin{pmatrix} 0 & 0 & 1 \\ 0 & 0 & 0 \\ -1 & 0 & 0 \end{pmatrix}. \tag{S8}$$

At this point we can already see why the s-polarized components of the reflected field do not affect the lateral force: because they do not have the appropriate parity in their plane wave decomposition to achieve an unbalanced force after the summation or integration of the individual forces of each of the plane wave components. Exactly the same can be said about the y-component of the source dipole $\mathbf{p}$. With such a big simplification, we may substitute eq. (S8) into eq. (S7), so that the relevant components of the reflected fields are given by:

$$\mathbf{E}_{\text{ref}}(k_x, k_y, z)\bigg|_{\substack{k_x \text{ odd} \\ k_y \text{ even}}} = \frac{i\omega^2 \mu_0 \mu_1}{8\pi^2} \cdot \frac{k_x r^p(k_x, k_y)}{k_1^2} \cdot \begin{pmatrix} -p_z \\ 0 \\ p_x \end{pmatrix} \cdot e^{ik_{z_1}(z+h)}. \tag{S9}$$

where $k_{z_1} = (k_1^2 - k_x^2 - k_y^2)^{1/2}$.

This reflected field can be substituted into the expression of the force [eqs. (S5) and (S6)], yielding an exact expression for the force:

$$\langle F_x \rangle = \iint\limits_{-\infty}^{+\infty} \frac{1}{2} \operatorname{Re}\left[ k_x^2 \cdot \frac{\omega^2 \mu_0 \mu_1}{8\pi^2} \frac{r^p(k_x, k_y)}{k_1^2} (p_x^* p_z - p_z^* p_x) \cdot e^{ik_{z_1}(2h)} \right] dk_x dk_y, \tag{S10}$$

We can work with this expression in order to simplify it

$$\langle F_x \rangle = \iint\limits_{-\infty}^{+\infty} \frac{1}{2} \operatorname{Re}\left[ k_x^2 \cdot \frac{\omega^2 \mu_0 \mu_1}{8\pi^2} \frac{r^p(k_x, k_y)}{k_1^2} (p_x^* p_z - p_z^* p_x) \cdot e^{ik_{z_1}(2h)} \right] dk_x dk_y, \quad \text{Apply } \frac{\omega^2 \mu_0 \mu_1}{k_1^2} = \frac{1}{\epsilon_0 \epsilon_1}.$$

$$= \iint\limits_{-\infty}^{+\infty} \frac{1}{2} \operatorname{Re}\left[ k_x^2 \cdot \frac{r^p(k_x, k_y)}{8\pi^2 \epsilon_0 \epsilon_1} (p_x^* p_z - p_z^* p_x) \cdot e^{ik_{z_1}(2h)} \right] dk_x dk_y, \qquad \text{Apply } [p_x^* p_z - p_z^* p_x] = 2i \operatorname{Im}[p_x^* p_z]$$

$$= \iint\limits_{-\infty}^{+\infty} \frac{1}{2} \operatorname{Re}\left[ 2i k_x^2 \cdot \frac{r^p(k_x, k_y)}{8\pi^2 \epsilon_0 \epsilon_1} \operatorname{Im}[p_x^* p_z] \cdot e^{ik_{z_1}(2h)} \right] dk_x dk_y, \qquad \text{Apply } \operatorname{Re}(iz) = -\operatorname{Im}(z)$$

$$= \iint\limits_{-\infty}^{+\infty} -\frac{1}{2} \operatorname{Im}\left[ 2 k_x^2 \cdot \frac{r^p(k_x, k_y)}{8\pi^2 \epsilon_0 \epsilon_1} \operatorname{Im}[p_x^* p_z] \cdot e^{ik_{z_1}(2h)} \right] dk_x dk_y, \qquad \text{Apply linearity}$$

we obtain the simplified expression:

$$\langle F_x \rangle = -\frac{1}{8\pi^2 \epsilon_0} \operatorname{Im}[p_x^* p_z] \iint\limits_{-\infty}^{+\infty} k_x^2 \operatorname{Im}\left[ \frac{1}{\epsilon_1} r^p(k_x, k_y) e^{ik_{z_1}(2h)} \right] dk_x dk_y, \tag{S11}$$

now we can write the integral in terms of polar wavevector coordinates by applying the following change of variables ($k_x = k_t \cos\alpha, k_y = k_t \sin\alpha$), so that the integral changes[1] according to:

$$\iint\limits_{-\infty}^{+\infty} f(k_x, k_y) \cdot dk_x dk_y = \int\limits_0^{+\infty} \int\limits_0^{2\pi} f(k_t \cos\alpha, k_t \sin\alpha) \cdot k_t \cdot d\alpha \cdot dk_t, \tag{S12}$$

which, applied to eq. (S11) gives:

$$\langle F_x \rangle = -\frac{1}{8\pi^2 \epsilon_0} \operatorname{Im}[p_x^* p_z] \int\limits_0^{+\infty} \int\limits_0^{2\pi} k_t^3 \cdot \cos^2(\alpha) \operatorname{Im}\left[ \frac{1}{\epsilon_1} r^p(k_t) e^{ik_{z_1}(2h)} \right] d\alpha \cdot dk_t, \tag{S13}$$

---

[1] Both $r^p(k_x, k_y) = r^p(k_t)$ and $k_{z_1} = k_{z_1}(k_t)$ are independent of $\alpha$.

S4

since everything else is independent of $\alpha$, we can evaluate the integral $\int_0^{2\pi} \cos^2(\alpha) d\alpha = \pi$, and we obtain:

$$\langle F_x \rangle = -\frac{1}{8\pi\epsilon_0} \text{Im}[p_x^* p_z] \int_0^{+\infty} k_t^3 \text{Im}\left[\frac{1}{\epsilon_1} r^p(k_t) e^{ik_{z_1}(2h)}\right] \cdot dk_t, \tag{S14}$$

**Normalization of parameters**

Equation (S14) above constitutes an exact expression for the lateral force, and could be used as Eq. (1) in the main text. However, working with the dipolar moments $p_x$ and $p_z$ is experimentally unintuitive. We would like to normalize the variables and introduce an experimentally intuitive measure of the amplitude of excitation of the dipole.

Firstly, we introduce the normalized transverse wave-vector $k_{tr} = k_t/k_0$, so that $dk_t = k_0 dk_{tr}$, and we can also write $k_{z1} = (k_1^2 - k_t^2)^{1/2} = k_0(n_1^2 - k_{tr}^2)^{1/2}$, where $n_1 = (\epsilon_1 \mu_1)^{1/2}$ is the effective index of medium 1, and $k_0 = 2\pi/\lambda$. Substituting all this into eq. (S14) gives:

$$\langle F_x \rangle = -\frac{k_0^4}{8\pi\epsilon_0} \text{Im}[p_x^* p_z] \int_0^{+\infty} k_{tr}^3 \text{Im}\left[\frac{1}{\epsilon_1} r^p(k_{tr}) e^{i2\pi \frac{2h}{\lambda} \sqrt{n_1^2 - k_{tr}^2}}\right] \cdot dk_{tr}, \tag{S15}$$

We will see that it is very convenient to introduce the power radiated by the dipole as a measure of its amplitude. In order to do this, we will from now on assume that medium 1 is free space ($\epsilon_1 = \mu_1 = n_1 = 1$). The power radiated by a dipole **p** in free space is given by:

$$P_{rad} = \frac{c_0 k_0^4}{12\pi\epsilon_0} |\mathbf{p}|^2$$

$$= \underbrace{\frac{c_0 k_0^4}{12\pi\epsilon_0}(|p_x|^2 + |p_z|^2)}_{P_{rad}^{xz}} + \frac{c_0 k_0^4}{12\pi\epsilon_0}|p_y|^2, \tag{S16}$$

so that $P_{rad}^{xz}$ can be substituted into eq. (S15) giving:

$$\langle F_x \rangle = -\frac{3}{2c_0} P_{rad}^{xz} \frac{\text{Im}[p_x^* p_z]}{|p_x|^2 + |p_z|^2} \int_0^{+\infty} k_{tr}^3 \text{Im}\left[r^p(k_{tr}) e^{i2\pi \frac{2h}{\lambda} \sqrt{1 - k_{tr}^2}}\right] \cdot dk_{tr}. \tag{S17}$$

Now we turn our attention to the polarization of the dipole. Let us denote the polarization vector of the dipole ignoring its y-component as $\mathbf{p_{xz}} = [p_x, 0, p_z]^T$. The "spin" of a dipole polarized in the $xz$ plane can be quantified using an expression similar to the third Stokes' parameter used to quantify the spin of a polarized electric field propagating along $y$.

$$\begin{aligned}\eta_{\text{pol}} &= 2\frac{(\text{Im}[\mathbf{p_{xz}}] \times \text{Re}[\mathbf{p_{xz}}]) \cdot \hat{\mathbf{y}}}{|\mathbf{p_{xz}}|^2} \\ &= 2\frac{\text{Im}[p_x p_z^*]}{|p_x|^2 + |p_z|^2} \\ &= \frac{|p_{\text{lcp}}|^2 - |p_{\text{rcp}}|^2}{|p_{\text{lcp}}|^2 + |p_{\text{rcp}}|^2} = \begin{cases} +1 & \text{for anti-clockwise (left-handed) circular polarization} \\ 0 & \text{for linear polarization} \\ -1 & \text{for clockwise (right-handed) circular polarization} \end{cases}\end{aligned} \tag{S18}$$

which, upon substitution into eq. (S17) yields the final expression used in the main text for the time-averaged lateral force:

$$\langle F_x \rangle = -\frac{3}{4c_0} P_{\text{rad}}^{xz} \eta_{\text{pol}} \int_0^{+\infty} k_{tr}^3 \, \text{Im}\left[ r^p(k_{tr}) e^{i 2\pi \frac{2h}{\lambda} \sqrt{1-k_{tr}^2}} \right] \cdot dk_{tr}. \qquad (S19)$$

## 2. Optimization of dielectric slab

Example 1 in the main text considers a dielectric slab with index $n_{\text{slab}}$ and thickness $t_{\text{slab}}$ placed on top of an infinite substrate with index $n_{\text{subs}}$. Such a structure will support guided TM modes in the slab, and will be suitable for achieving lateral forces on nearby circularly polarized dipoles.

The index ($n_{\text{slab}}$) and thickness ($t_{\text{slab}}$) of the slab will determine the existence of TM modes, as well as their effective index and excitation amplitudes, and therefore the recoil forces. In this section, we perform a numerical study in order to optimize the thickness of the slab, considering a substrate of $n_{\text{subs}} = 1.45$. We numerically apply eqs. (S19) and (A8) for different values of $n_{\text{slab}}$ and $t_{\text{slab}}$. The resulting force map at $h = 0.1\lambda$, valid at any frequency, is shown in Fig. S1.

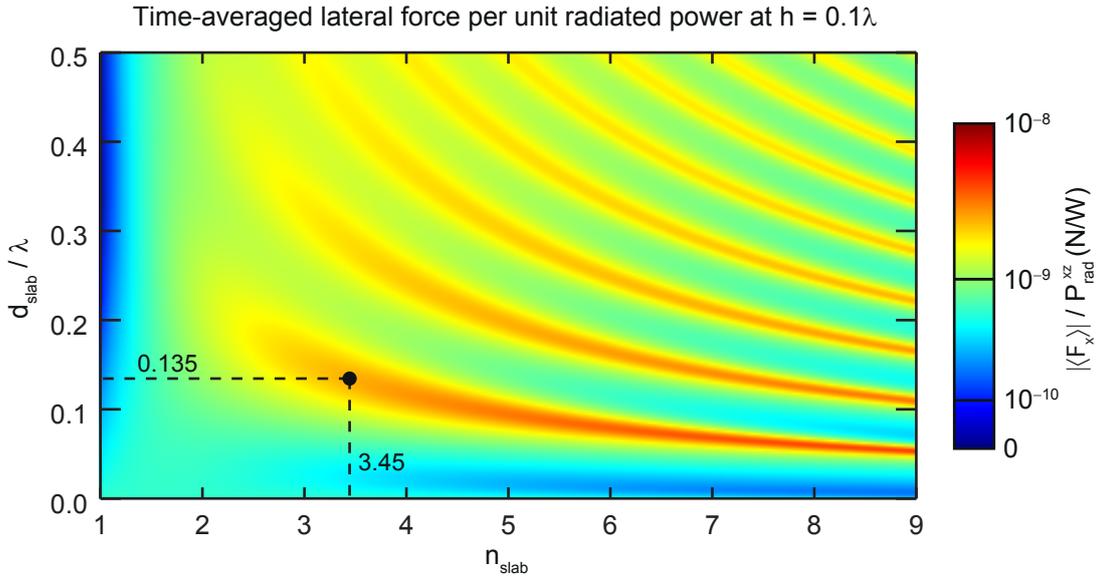

**Figure S1.** Numerical calculation of the lateral force acting on a dipole over a dielectric slab. Fixed parameters considered for this calculation were $h = 0.1\lambda$, $\eta_{\text{pol}} = 1$, and a silica substrate $n_{\text{subs}} = 1.45$.

## 3. Derivation of the different terms

### Motivation for the separation into terms

The final equation for the lateral force [eq. (S19)] is straight forward to compute numerically, but carries a limited physical insight. By looking at the integrand (see Fig. S2), we can understand that the force may come from several contributions.

S6

- Firstly, we note that the reflection coefficient $r^p(k_{tr})$ has a resonant behavior whenever there is a TM polarized mode supported by the surface/slab/waveguide. This will provide peaks in the integrand of eq. (S19) and therefore will contribute to the force.

- On the other hand, even if $r^p$ does not have resonances, the term $k_{tr}^3 \cdot \exp(4\pi h(1 - k_{tr}^2)^{1/2})$ will initially grow as $k_{tr}$ grows, but will eventually decay due to the exponential term, thus a bell-shaped curve will occur in the integrand, of higher amplitude the closer the dipole is to the surface, and will also contribute to the force.

Both contributions can be clearly seen in Fig. S2, showing a representation of the exact curve of the integrand as a function of $k_{tr}$ for a given height and geometry, and also depicting the two contributions. Finally, when the distance $h$ is increased into the far-field, all the components of the integrand that are above $k_{tr} > 1$ vanish, together with the two terms. We are only left with the force caused by the propagating plane waves, which we define as the third term.

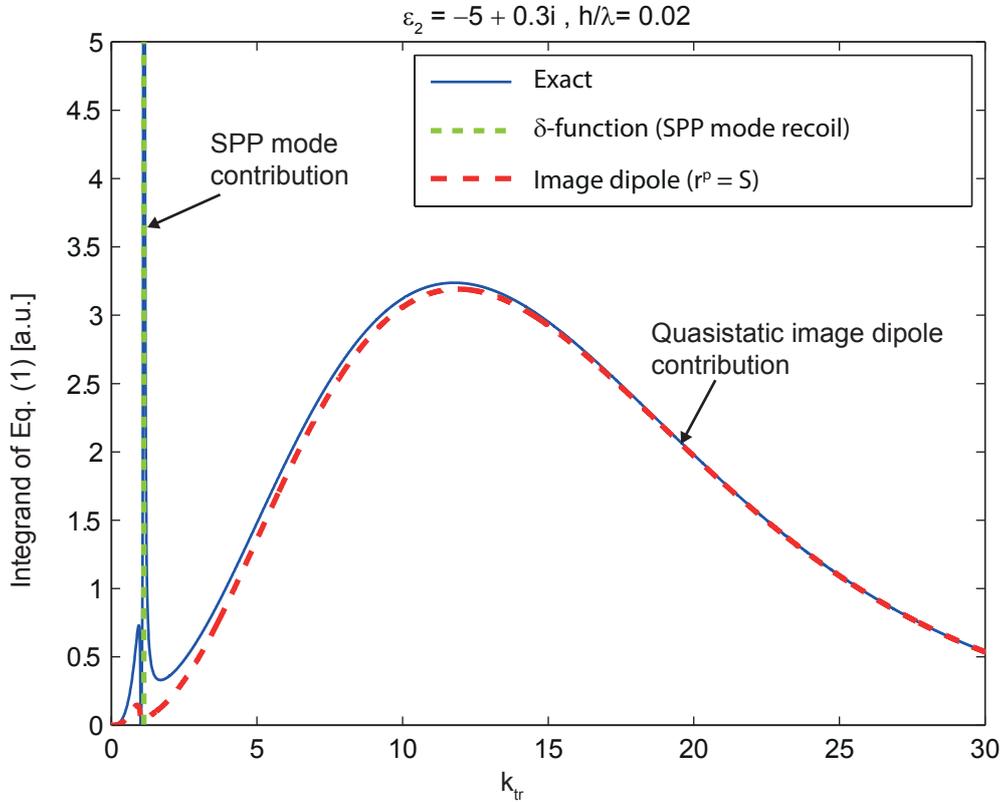

**Figure S2.** Integrand of eq. (S19) for a single metal substrate with $\epsilon_2 = -5 + 0.3i$, supporting an SPP mode and an out of phase image dipole. In this case the dipole is placed at a distance $h = 0.02$.

**Mode recoil term**

The reflection coefficient $r^p(k_{tr})$, according to eqs. (A7) and (A8), exhibits resonant peaks at the values of $k_{tr} = n_{\text{eff},1}, n_{\text{eff},2}, \ldots$ where $n_{\text{eff},k}$ is the effective index of the k-th TM guided mode of the surface. Since $r^p(k_{tr})$ appears in the integrand of eq. (S19), the peaks will contribute to the force. The higher the mode excitation, the higher this contribution will be. We interpret this as the recoil force due to the directional excitation of the guided mode. It is known that any TM mode supported by the surface will be excited directionally by a circularly polarized dipole, as discussed in the main text.

To find an approximate expression for this term, we simply approximate the imaginary part of $r^p(k_{tr})$ as a series of $\delta$-functions corresponding to the different TM modes supported by the waveguide, labelled from $k = 1...N$ (N can be zero if there are no supported modes, so this term vanishes):

$$\text{Im}[r^p(k_{tr})] \approx \sum_{k=1}^{N} R_k \delta(k_{tr} - n_{\text{eff},k}) \tag{S20}$$

where $R_k$ is the area under the resonant peak in $r^p$, as defined in the main text, and $n_{\text{eff},k}$ is the effective index of the k-th mode.

Thanks to the sieving property of the $\delta$ function $\int \delta(x - x_0) f(x) dx = f(x_0)$, it is trivial to substitute eq. (S20) into eq. (S19) and perform the integration, resulting in Eq. 3 in the main text.

**Image dipole term**

For the image dipole term, we apply the quasistatic approximation to the reflection coefficient. In the quasistatic limit, the z-component of the wave-vector is given by $k_{zi} = (n_i k_0^2 - k_{tr}^2)^{1/2} \xrightarrow{k_{tr} \gg k_0} k_{zi} \approx i k_{tr}$, which when substituted into the equations of the reflection coefficient [eqs. (A7) and (A8)] result in a constant value for $r^p$. Therefore, we can take $r^p$ to be constant and equal to the complex image coefficient $S$:

$$r^p(k_{tr}) \approx S = \lim_{k_{tr} \to \infty} r^p(k_{tr}) = \frac{\epsilon_2 - \epsilon_1}{\epsilon_2 + \epsilon_1} \tag{S21}$$

where $\epsilon_1$ and $\epsilon_2$ are the relative permittivities of the medium above and below the $z = 0$ surface, respectively. If $r^p$ is approximated by the constant $S$, it can be taken out of the integral in eq. (S19), yielding:

$$\langle F_x \rangle = -\frac{3}{4c_0} P_{\text{rad}}^{xz} \eta_{\text{pol}} \text{Im} \left[ S \int_0^{+\infty} k_{tr}^3 e^{i2\pi \frac{2h}{\lambda} \sqrt{1-k_{tr}^2}} \cdot dk_{tr} \right]. \tag{S22}$$

and the integral now has an analytical solution, as follows:

$$\text{Im} \left[ S \int_0^{+\infty} k_{tr}^3 e^{i2\pi \frac{2h}{\lambda} \sqrt{1-k_{tr}^2}} \cdot dk_{tr} \right]$$
$$= \frac{3 \, \text{Im}[S] \cos(4\pi h/\lambda)}{128\pi^4} \left(\frac{h}{\lambda}\right)^{-4} + \frac{3 \, \text{Im}[S] \sin(4\pi h/\lambda)}{32\pi^3} \left(\frac{h}{\lambda}\right)^{-3} - \frac{\text{Im}[S] \cos(4\pi h/\lambda)}{8\pi^2} \left(\frac{h}{\lambda}\right)^{-2} \tag{S23}$$
$$+ \frac{3 \, \text{Re}[S] \sin(4\pi h/\lambda)}{128\pi^4} \left(\frac{h}{\lambda}\right)^{-4} - \frac{3 \, \text{Re}[S] \cos(4\pi h/\lambda)}{32\pi^3} \left(\frac{h}{\lambda}\right)^{-3} - \frac{\text{Re}[S] \sin(4\pi h/\lambda)}{8\pi^2} \left(\frac{h}{\lambda}\right)^{-2}.$$

The quasistatic approximation is only valid when the electric dipole source is very close to the surface, in the limit $h \to 0$, therefore we can make a Taylor expansion around $h = 0$ of the sine and cosine functions in eq. (S23) and keep only the terms with the most negative exponent of $h$, which will dominate over the others at low heights. This gives:

$$\text{Im}\left[S\int_0^{+\infty} k_{tr}^3 e^{i4\pi \frac{h}{\lambda}\sqrt{1-k_{tr}^2}} \cdot dk_{tr}\right] \approx \frac{3\,\text{Im}[S]}{128\pi^4}\left(\frac{h}{\lambda}\right)^{-4} + \text{higher order terms...} \tag{S24}$$

Substituting eq. (S24) into eq. (S22) we arrive at:

$$\langle F_x \rangle \approx -\frac{3}{4c_0} P_{\text{rad}}^{xz} \eta_{\text{pol}} \frac{3}{128\pi^4} \text{Im}[S]\left(\frac{h}{\lambda}\right)^{-4}. \tag{S25}$$

which is the final form of Eq. (4) in the main text.

Also notice that, by assuming $r^p \approx S$, the reflected fields of the substrate [eqs. (A5) to (A6b) in the Annex A] are exactly equivalent to the fields created by a dipole [eqs. (A1) to (A3b)] whose dipole moment corresponds to associating each charge $q$ of the source dipole to a mirrored image charge $q' = -Sq$, resulting in an image dipole with polarization $\mathbf{p}_{\text{image}} = [-Sp_x, -Sp_y, Sp_z]$ located at $z = -h$. This is consistent with image-theory of static charges over a substrate. For this reason, this term represents the contribution to the force of the quasistatic image dipole.

**Lateral force between two arbitrary dipoles**

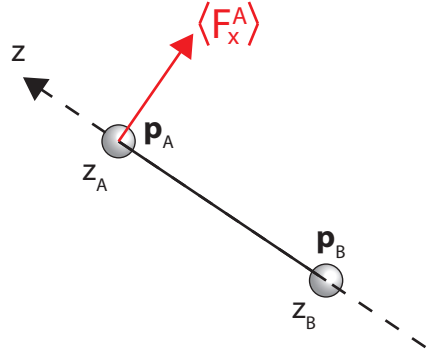

**Figure S3.** Depiction of the lateral force acting between two ideal point dipoles.

We can generalise the previous result to the case of two arbitrarily polarized dipoles. Consider two dipoles with polarization $\mathbf{p_A}$ and $\mathbf{p_B}$ separated a distance $d$. Without loss of generality, we can assume dipole A to be located at $(x, y, z) = (0, 0, z_A)$ and dipole B at $(x, y, z) = (0, 0, z_B)$, as shown in Fig. S3. We can calculate the force produced by dipole B on dipole A. The fields of dipole B can be obtained by using eqs. (A1) to (A3b), and substituted into the expression of the force eq. (S1). Following the same steps we did for the dipole over the substrate, we arrive at a final expression for the lateral force ($x$ component) exerted between the two dipoles:

$$\langle F_x^A \rangle = \pm\frac{3}{4c_0}\sqrt{P_{\text{rad},xz}^A P_{\text{rad},xz}^B} \cdot \text{Re}\left[\frac{p_x^{A*}p_z^B + p_z^{A*}p_x^B}{|\mathbf{p_{xz}^A}|\cdot|\mathbf{p_{xz}^B}|} \cdot \int_0^{+\infty} k_{tr}^3 e^{ik_z|z_A-z_B|} \cdot dk_{tr}\right]. \tag{S26}$$

where the upper sign is used if $z_A > z_B$ and viceversa, and $\mathbf{p_{xz}^{A,B}}$ corresponds to the dipole moment of dipoles A or B neglecting the y-component. It is surprising that such a two point system can actually exhibit a lateral force between the dipoles, orthogonal to the line joining

S9

them, thanks to their polarization.

If we make dipole B equal to an image dipole $\mathbf{p_B} = -S \cdot [p_x^A, p_y^A, -p_z^A]$, as quasistatic image theory requires, then we have:

$$\sqrt{P_{\text{rad},xz}^A P_{\text{rad},xz}^B} = |S| P_{\text{rad},xz}^A,$$
$$|\mathbf{p_{xz}^A}| \cdot |\mathbf{p_{xz}^B}| = |S| |\mathbf{p_{xz}^A}|^2, \text{ and}$$
$$p_x^{A*} p_z^B + p_z^{A*} p_x^B = S(p_x^{A*} p_z^A - p_z^{A*} p_x^A) = 2iS \, \text{Im}[p_x^{A*} p_z^A] = iS \cdot \eta_{\text{pol}} \cdot |\mathbf{p_{xz}^A}|^2,$$

which makes eq. (S26) identical to eq. (S22), as expected.

## 4. Numerical simulations

In order to cross-check the results presented in the main text, we performed a time-domain numerical calculation of the electromagnetic fields and the associated optical forces in the two scenarios described in the main text. We used the commercial simulation software CST Microwave Studio.

We simulated the circularly polarized point dipole by using two orthogonal subwavelength discrete current ports, excited $\pi/2$ out of phase to each other, placed at a variable distance $h$ above a surface. The boundary conditions in all directions were set to perfectly matched layers (PMLs). The surface at $z = 0$ was a dielectric slab over a dielectric substrate, in the first example, or a metallic substrate, in the second example, as described in the main text.

After obtaining the electromagnetic fields created by the dipole, we computed Maxwell's stress tensor in the space surrounding the dipole, and integrated it in a box surrounding the dipole, to obtain the total average force acting on it [1]. We repeated this procedure for different values of the distance $h$. To make sure that our simulation results of the force were consistent, we made sure that changing the size of the box of integration resulted in the same value of the force, since the force calculated by integrating Maxwell's stress tensor on a box depends only on the objects contained within it, and not on the size or shape of the box [1].

Also, we noticed that the size of the simulation region affected our results. It is well known that PML boundary conditions are not perfect, and they always show some unwanted reflection. Ideally, the PML layers should be as far from the dipole as possible for best results, requiring longer simulation times. As shown in Fig. S4, the numerical calculation of the force approaches the analytical calculation as the size of the simulation region ($D$) is increased.

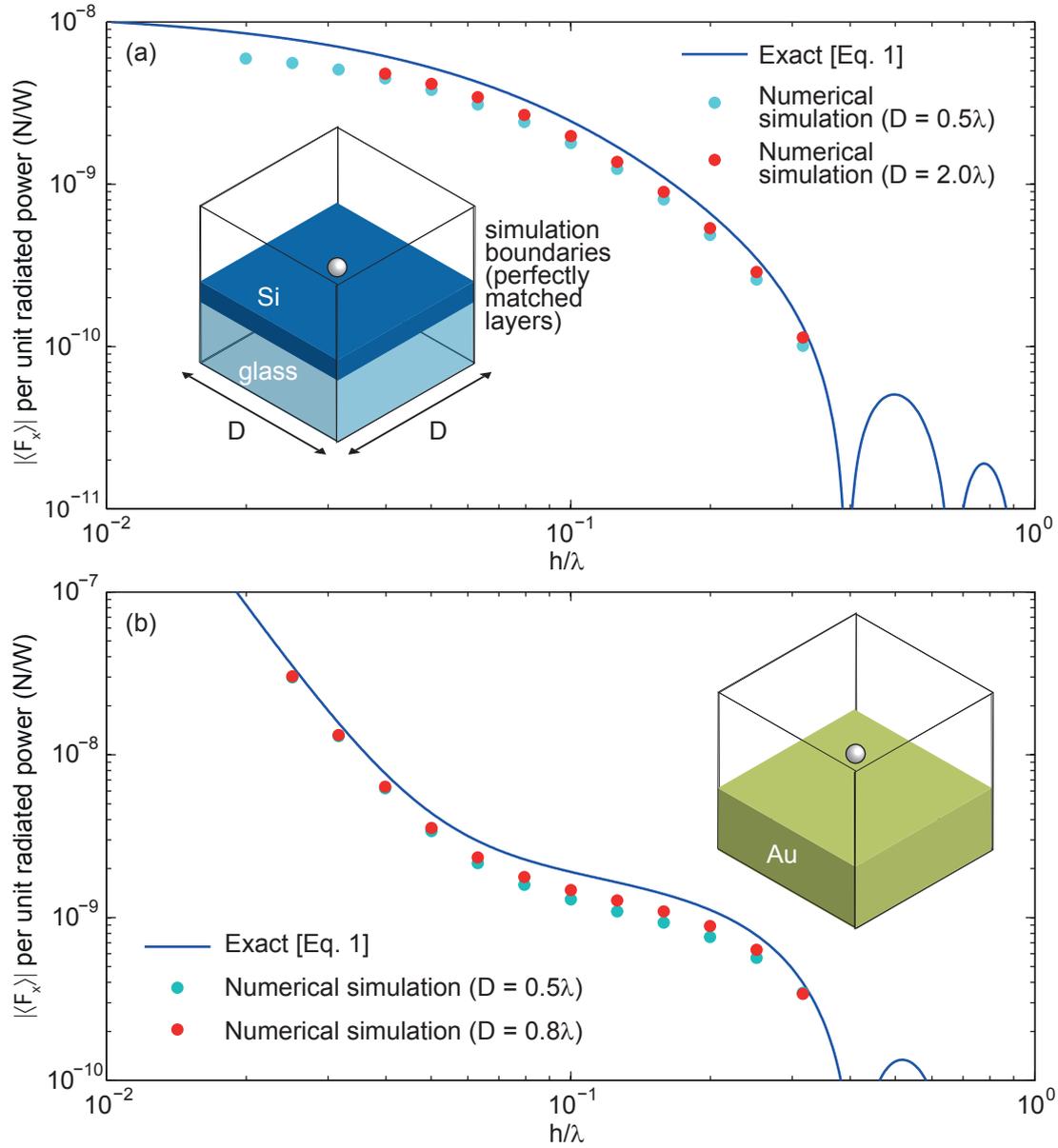

**Figure S4.** Numerical calculation of the lateral force acting on a dipole using the Maxwell's stress tensor for a dipole over (a) a dielectric slab, or (b) a metal substrate, corresponding to the same parameters used by the two examples in the main text. Simulations with different simulation region sizes are included, showing that the numerical force approaches the exact analytical expression as the simulation size is increased.

## 5. Practical force calculations

Consider a spherical gold nanoparticle at a height $h$ above a surface, being illuminated by an unfocused laser beam (which can locally be considered as a plane wave) at a grazing angle (with its electric field contained within the $x - z$ plane). The incident power density is given by the power of the laser divided by the cross-sectional area of the beam :

$$W_{\text{inc}} = \frac{P_{\text{laser}}}{A_{\text{illumination}}}. \quad (S27)$$

which, for a typical 10 mW laser beam illuminating a spot of radius 10 $\mu$m is equal to $W_{\text{inc}} = 3.18 \times 10^{-11}$ W/nm$^2$.

The gold nanoparticle, if it has subwavelength dimensions (radius $R \ll \lambda$), can be modelled by a polarizable particle with isotropic polarizability $\alpha$ given by the Rayleigh/dipole approximation:

$$\alpha = 4\pi R^3 \epsilon_0 \epsilon_m \left( \frac{\epsilon_p - \epsilon_m}{\epsilon_p + 2\epsilon_m} \right), \quad (S28)$$

where $\epsilon_0$ is the permittivity of free space, $\epsilon_m$ is the relative permittivity of the surrounding medium, and $\epsilon_p$ is the relative permittivity of the particle. For the gold particle we take the values of Johnson and Christy [46]. The particle will get polarized by the incident plane wave electric field according to $\mathbf{p} = \alpha \mathbf{E}_{\text{ind}}$.

The scattered power of the particle will be given by $P_{\text{sca}} = \sigma_{\text{sca}} W_{\text{inc}}$; Similarly, the absorbed power of the particle will be $P_{\text{abs}} = \sigma_{\text{abs}} W_{\text{inc}}$. The scattering and absorption cross-sections for a polarizable particle are given by:

$$\sigma_{\text{sca}} = |\alpha|^2 \frac{k_0^4}{6\pi \epsilon_0^2 \epsilon_m^2},$$
$$\sigma_{\text{abs}} = \text{Im}[\alpha] \frac{k_0}{\epsilon_0 \epsilon_m}. \quad (S29)$$

The scattering and absorption cross-sections for gold particles of different radii are plotted in Fig S5. We see that at the localized plasmonic resonance ($\lambda = 520$ nm) the scattering cross section reaches a value of $\sigma_{\text{sca}} = 1100$ nm$^2$ for a gold nanoparticle of radius $R = 30$ nm in vacuum. Therefore, the power scattered by the particle at that wavelength is $P_{\text{sca}} = \sigma_{\text{sca}} W_{\text{inc}} = 3.5 \times 10^{-8}$ W.

The lateral force acting on the particle due to the scattered power can be calculated by applying Eq. (1) in the main text, taking $P_{\text{rad}}^{xy} = P_{\text{sca}}$, and taking the polarization factor $\eta_{\text{pol}}$ equal to that of the incoming plane wave (in practice, an anisotropic polarizability could introduce a difference between the polarization of the dipole and that of the incoming plane wave, but we consider the polarizability to be isotropic as a good approximation). The graph of the normalized force should therefore be multiplied by this scattered power to obtain the force in Newtons.

The illuminating plane wave will also exert a force on the particle due to the transfer of its linear momentum. This pressure force will be directed in the direction of illumination, and is given by (see ref. [1]):

$$\langle F_{\text{pressure}} \rangle = n_m P_{\text{abs}}/c_0, \quad (S30)$$

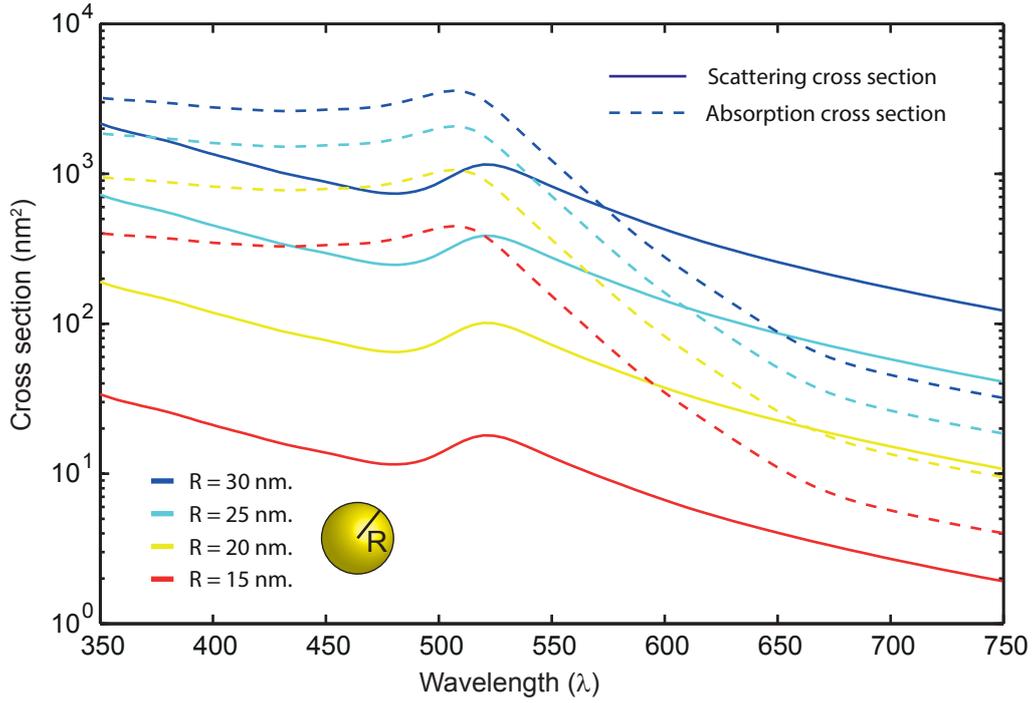

**Figure S5.** Scattering and absorption cross section for Au particles of different radii, under the Rayleigh approximation.

where $n_m$ is the refractive index of the medium and $c_0$ is the speed of light. In the case of the $R = 30$ nm gold particle in vacuum under the illumination discussed previously at the same wavelength $\lambda = 520$ nm, the pressure force is equal to $3.3 \times 10^{-16}$ N.

So, in this particular situation, the lateral force and the pressure force are of comparable magnitude and directed in orthogonal directions. In general, the ratio of magnitudes of the lateral force to the pressure force is proportional to:

$$\frac{\langle F_x \rangle}{\langle F_{\text{pressure}} \rangle} \propto \frac{\sigma_{\text{sca}}}{\sigma_{\text{abs}}} \propto \frac{k_0^3 |\alpha|^2}{\text{Im}[\alpha]} \propto \left(\frac{R}{\lambda}\right)^3. \tag{S31}$$

Additional forces in the system will be caused by the reflection of the incident plane wave on the substrate, introducing an additional pressure force and gradient force due to the standing wave that is formed upon such reflection. Both will depend on the absorption of the particle and therefore will also follow the ratio of eq. (S31) with the lateral force. In addition, the reflected wave could also polarize the dipole. Calculating the specific values and effects of this would be necessary considerations in a practical situation, but is beyond the scope of the present study.

## Annex A: Spatial decomposition of dipole fields and reflected fields

The fields of the source dipole can be written as a superposition of plane waves and evanescent waves with different transverse wave-vectors $(k_x, k_y)$ for both s-polarized and p-polarized plane waves, as:

$$\mathbf{E}_{\text{dip}}(\mathbf{r}) = \iint_{-\infty}^{+\infty} \overbrace{\mathbf{E}_{\text{dip}}(k_x, k_y) \cdot e^{ik_{z_1}|z-h|}}^{\mathbf{E}_{\text{dip}}(k_x,k_y,z)} \cdot e^{ik_x x + ik_y y} dk_x dk_y,$$

$$\text{where } \mathbf{E}_{\text{dip}}(k_x, k_y) = \underbrace{\mathbf{E}^{\mathbf{s}}_{\text{dip}}(k_x, k_y)}_{\text{s-polarized}} + \underbrace{\mathbf{E}^{\mathbf{p}}_{\text{dip}}(k_x, k_y)}_{\text{p-polarized}}$$

(A1)

For a dipole source $\mathbf{p} = [p_x, p_y, p_z]^T$, the spatial spectrum of the electric field can be deduced by applying the Weyl mathematical identity to the well-known expression for the fields of a dipole in free space, and is given by (see Ref. [1]):

$$\mathbf{E_{dip}}(k_x, k_y) = \frac{i\omega^2 \mu_0 \mu_1}{8\pi^2} \left( \overset{\leftrightarrow}{\mathbf{M}} \mathbf{p} \right),$$

$$\text{where } \overset{\leftrightarrow}{\mathbf{M}} = \overset{\leftrightarrow}{\mathbf{M}^{\mathbf{s}}}(k_x, k_y) + \overset{\leftrightarrow}{\mathbf{M}^{\mathbf{p}}}(k_x, k_y)$$

(A2)

where the tensor matrices $\overset{\leftrightarrow}{\mathbf{M}^{\mathbf{s}}}$ and $\overset{\leftrightarrow}{\mathbf{M}^{\mathbf{p}}}$ are:

$$\overset{\leftrightarrow}{\mathbf{M}^{\mathbf{s}}}(k_x, k_y) = \frac{1}{k_{z_1}(k_x^2 + k_y^2)} \begin{pmatrix} k_y^2 & -k_x k_y & 0 \\ -k_x k_y & k_x^2 & 0 \\ 0 & 0 & 0 \end{pmatrix}$$

(A3a)

$$\overset{\leftrightarrow}{\mathbf{M}^{\mathbf{p}}}(k_x, k_y) = \frac{1}{k_1^2(k_x^2 + k_y^2)} \begin{pmatrix} k_x^2 k_{z_1} & k_x k_y k_{z_1} & \mp k_x(k_x^2 + k_y^2) \\ k_x k_y k_{z_1} & k_y^2 k_{z_1} & \mp k_y(k_x^2 + k_y^2) \\ \mp k_x(k_x^2 + k_y^2) & \mp k_y(k_x^2 + k_y^2) & (k_x^2 + k_y^2)^2/k_{z_1} \end{pmatrix},$$

(A3b)

and the upper sign is used for $z > h$ while the lower sign is used for $z < h$.

Knowing the fields generated by the dipole (often called the primary fields) we can now calculate the reflected fields by the surface, simply by considering that each plane wave component[2] will be reflected according to its polarization and its transverse wavevector $k_t$ with the appropriate Fresnel reflection coefficient $r^p(k_t)$ or $r^s(k_t)$. Doing this yields the spatial spectrum of the reflected fields:

$$\mathbf{E}_{\text{ref}}(\mathbf{r}) = \iint_{-\infty}^{+\infty} \overbrace{\mathbf{E}_{\text{ref}}(k_x, k_y) \cdot e^{ik_{z_1}(z+h)}}^{\mathbf{E}_{\text{ref}}(k_x,k_y,z)} \cdot e^{ik_x x + ik_y y} dk_x dk_y,$$

$$\text{where } \mathbf{E}_{\text{ref}}(k_x, k_y) = \underbrace{\mathbf{E}^{\mathbf{s}}_{\text{ref}}(k_x, k_y)}_{\text{s-polarized}} + \underbrace{\mathbf{E}^{\mathbf{p}}_{\text{ref}}(k_x, k_y)}_{\text{p-polarized}}$$

(A4)

where, similar to what we did previously

---

[2] Mathematically this also works for the evanescent field components, which have the same analytical form as plane waves simply by taking $k_t > k_1$ which makes $k_{z_1}$ imaginary.

$$\mathbf{E_{ref}}(k_x, k_y) = \frac{i\omega^2 \mu_0 \mu_1}{8\pi^2} \left( \overset{\leftrightarrow}{\mathbf{M}}_{\text{ref}} \, \mathbf{p} \right), \tag{A5}$$

where $\overset{\leftrightarrow}{\mathbf{M}}_{\text{ref}} = \overset{\leftrightarrow}{\mathbf{M}}^{\mathbf{s}}_{\text{ref}}(k_x, k_y) + \overset{\leftrightarrow}{\mathbf{M}}^{\mathbf{p}}_{\text{ref}}(k_x, k_y)$

and the tensors $\overset{\leftrightarrow}{\mathbf{M}}^{\mathbf{s}}_{\text{ref}}(k_x, k_y)$ and $\overset{\leftrightarrow}{\mathbf{M}}^{\mathbf{p}}_{\text{ref}}(k_x, k_y)$ are given by:

$$\overset{\leftrightarrow}{\mathbf{M}}^{\mathbf{s}}_{\text{ref}}(k_x, k_y) = \frac{r^s(k_x, k_y)}{k_{z_1}(k_x^2 + k_y^2)} \begin{pmatrix} k_y^2 & -k_x k_y & 0 \\ -k_x k_y & k_x^2 & 0 \\ 0 & 0 & 0 \end{pmatrix} \tag{A6a}$$

$$\overset{\leftrightarrow}{\mathbf{M}}^{\mathbf{p}}_{\text{ref}}(k_x, k_y) = \frac{-r^p(k_x, k_y)}{k_1^2(k_x^2 + k_y^2)} \begin{pmatrix} k_x^2 k_{z_1} & k_x k_y k_{z_1} & +k_x(k_x^2 + k_y^2) \\ k_x k_y k_{z_1} & k_y^2 k_{z_1} & +k_y(k_x^2 + k_y^2) \\ -k_x(k_x^2 + k_y^2) & -k_y(k_x^2 + k_y^2) & -(k_x^2 + k_y^2)^2 / k_{z_1} \end{pmatrix}. \tag{A6b}$$

These reflected fields are often called the secondary fields of the dipole, and are the ones responsible of exerting a force on it. All these expressions can be found in Ref. [1].

For a single interface (upper medium with relative perittivity $\epsilon_i$, and lower medium with relative permittivity $\epsilon_j$), the reflection coefficient is given by:

$$r^p(k_x, k_y) = r^p(k_t) = \frac{\epsilon_2 k_{z1} - \epsilon_1 k_{z2}}{\epsilon_2 k_{z1} + \epsilon_1 k_{z2}}, \tag{A7}$$

where $k_{zi} = (k_i^2 - k_t^2)^{1/2} = k_0(n_i^2 - k_{tr}^2)^{1/2}$. For a single slab of permittivity $\epsilon_2$ and thickness $t$ over a substrate of permittivity $\epsilon_3$, the p-polarized reflection coefficient is given by:

$$r^p(k_x, k_y) = r^p(k_t) = \frac{r^p_{1,2} + r^p_{2,3} e^{2ik_{z2}t}}{1 + r^p_{1,2} r^p_{2,3} e^{2ik_{z2}t}}, \tag{A8}$$

where $r^p_{i,j}$ correspond to a single interface between medium $\epsilon_i$ and $\epsilon_j$ taken from eq. (A7).